\documentclass[11pt]{article}
\usepackage{amssymb,amsmath,amsfonts}
\usepackage{graphicx}
\usepackage{graphics}
\usepackage{epsfig}

\begin{document}

\title{Fermionic vacuum stresses in models with toroidal compact dimensions}
\author{A. A. Saharian$^{1,2}$\thanks{%
E-mail: saharian@ysu.am},\thinspace\ R. M. Avagyan$^{1,2}$,\thinspace\ G. H.
Harutyunyan$^{1}$,\thinspace\ G. H. Nikoghosyan$^{1}$ \\
\\
\textit{$^1$Institute of Physics, Yerevan State University,}\\
\textit{1 Alex Manoogian Street, 0025 Yerevan, Armenia} \vspace{0.3cm}\\
\textit{$^2$Institute of Applied Problems of Physics NAS RA,}\\
\textit{25 Hrachya Nersissyan Street, 0014 Yerevan, Armenia}}
\maketitle

\begin{abstract}
We investigate vacuum expectation value of the energy-momentum tensor for a
massive Dirac field in flat spacetime with a toroidal subspace of a general
dimension. Quasiperiodicity conditions with arbitrary phases are imposed on
the field operator along compact dimensions. These phases are interpreted in
terms of magnetic fluxes enclosed by compact dimensions. The equation of
state in the uncompact subspace is of the cosmological constant type. It is
shown that, in addition to the diagonal components, the vacuum
energy-momentum tensor has nonzero off-diagonal components. In special cases
of twisted (antiperiodic) and untwisted (periodic) fields the off diagonal
components vanish. For untwisted fields the vacuum energy density is
positive and the energy-momentum tensor obeys the strong energy condition.
For general values of the phases in the periodicity conditions the energy
density and stresses can be either positive or negative. The numerical
results are given for a Kaluza-Klein type model with two extra dimensions.
\end{abstract}

\bigskip


\bigskip

\section{Introduction}

The field theoretical models in background spacetimes with compact
dimensions appear in a number of theories in fundamental physics like string
theories, supergravities and Kaluza-Klein theories. The quantum creation of
universe with a compact space has been considered in \cite{Zeld84}-\cite%
{Lind04}. In this type of models the probability of inflation in the early
stages of the universe expansion is not exponentially small. The effects
caused by the non-trivial topology of the universe on cosmological scales
are discussed, for example, in \cite{Lach95,Levi02}. They include the ghost
images of galaxies and quasars, cosmological magnetic fields and observable
effects on cosmic microwave background. Physical models formulated on
background geometries with nontrivial topology also appear in a number of
condensed matter physics systems. Examples are topological structures of
graphene, like carbon nanotubes and nanoloops. The long wavelength
excitations of the electronic subsystem in those structures are described by
an effective field theory (Dirac model, see \cite{Gusy07,Cast09}) with
2-dimensional spatial topologies $R^{1}\times S^{1}$ and $T^{2}=S^{1}\times
S^{1}$, respectively.

In quantum field theory the nontrivial spatial topology is a source of a
number of interesting effects. In particular, the periodicity conditions
along compact dimensions modify the spectrum of quantum fluctuations of
fields and, as a consequence, the expectation values of the physical
characteristics are shifted by an amount that depends on the geometry and
topology of the compact subspace. This general phenomenon is known as the
topological Casimir effect (see \cite{Most97}-\cite{Dalv11}). The vacuum
energy in the topological Casimir effect depends on the size of compact
dimensions and this provides a stabilization mechanism for the corresponding
moduli fields. The topological Casimir energy may also appear as the source
of the accelerated expansion of uncompact subspace playing the role of the
dark energy at recent epoch of the Universe expansion (see, for example,
\cite{Milt03}-\cite{Wong15}).

An important physical characteristic for charged fields is the expectation
values of the current density. For a relatively simple model of toroidal
compactification in flat spacetime (for quantum field theory in models with
toroidal spatial dimensions see, for example, \cite{Khan14}), in references
\cite{Bell10}-\cite{Bell15} it has been shown that the nontrivial phases in
the periodicity conditions along compact dimensions give rise to nonzero
currents along those dimensions. The phases can be interpreted in terms of
magnetic fluxes enclosed by those dimensions. The currents in the compact
subspace are sources of magnetic fields having components along
uncompactified dimensions. The dependence of the vacuum energy density and
diagonal stresses for a massive fermionic field in the same model of flat
spacetime with a part of spatial dimensions compactified on a torus has been
studied in \cite{Bell09}. In the present paper we show that, in addition to
the diagonal components, the vacuum expectation value (VEV) of the
energy-momentum tensor may have nonzero off-diagonal components (vacuum
stresses) in the compact subspace.

The paper is organized as follows. In the next section present the problem
setup and the eigenmodes for a Dirac field obeying the quasiperiodicity
conditions along compact dimensions. The general formulas for the vacuum
energy density and stresses are obtained in section \ref{sec:EMT}. The
asymptotic and numerical analysis of the VEVs in a model with two compact
dimensions is presented in section \ref{sec:Asnum}. The main results are
summarized in section \ref{sec:Conc}.

\section{Problem setup}

\label{sec:Setup}

The background geometry we are going to consider is a flat spacetime with
topology $M^{p+1}\times T^{q}$, where $M^{p+1}$ is $(p+1)$-dimensional
Minkowski spacetime covered by the Cartesian coordinates $(z^{0}=t,\mathbf{z}%
_{p})=(z^{0},z^{1},\ldots ,z^{p})$, and $T^{q}=(S^{1})^{q}$ is a $q$%
-dimensional torus with the coordinates $\mathbf{z}_{q}=(z^{p+1},\ldots
,z^{D})$. The length of the $l$th compact dimension will be denoted by $%
L_{l} $ and, hence, $0\leq z^{l}\leq L_{l}$, $l=p+1,\ldots ,D$. The volume
of the compact subspace is expressed as $V_{q}=L_{p+1}\cdots L_{D}$. For the
uncompact dimensions, as usual, we have $z^{l}\in (-\infty ,+\infty )$. The
line element has the standard Minkowskian form
\begin{equation}
ds^{2}=\eta _{\mu \nu }dx^{\mu }dx^{\nu }=dt^{2}-d\mathbf{z}^{2},\;\mathbf{z}%
=(\mathbf{z}_{p},\mathbf{z}_{q}),  \label{ds}
\end{equation}%
and $\eta _{\mu \nu }$ is the Minkowski metric tensor in Cartesian
coordinates.

We are interested in the vacuum stresses for a massive Dirac field $\psi (x)$%
, $x=(t,\mathbf{z})$, induced by compactification of a part of spatial
dimensions. The field equation reads
\begin{equation}
\left( i\gamma ^{\mu }\partial _{\mu }-m\right) \psi (x)=0,  \label{Deq}
\end{equation}%
where $\gamma ^{\mu }$, $\mu =0,1,\ldots ,D$, are $N\times N$ Dirac matrices
with $N$ given by $N=2^{\left[ \frac{D+1}{2}\right] }$ and $[a]$ stands for
the integer part of $a$. The background geoemtry has non-trivial topology
and in order to fix the dynamics uniquely the periodicity conditions along
the compact dimensions have to be specified for the field operator. We
impose quasiperiodicity conditions%
\begin{equation}
\psi (t,\mathbf{z}_{p},\ldots ,z^{l}+L_{l},\ldots ,z^{D})=e^{i\alpha
_{l}}\psi (t,\mathbf{z}_{p},\ldots ,z^{l},\ldots ,z^{D}),  \label{Qper}
\end{equation}%
with phases $\alpha _{l}=\mathrm{const}$ and $l=p+1,\ldots ,D$. The special
cases of periodic and antiperiodic fields correspond to $\alpha _{l}=0$ and $%
\alpha _{l}=\pi $ (untwisted and twisted fields, respectively).

The VEV of the energy-momentum tensor $T_{\mu \nu }$ for the fermionic field
is expressed in terms of the mode sum over a complete set of normal modes $%
\psi _{\beta }^{(\pm )}(x)$, where $\beta $ is the set of quantum numbers
specifying the solutions to the field equation and upper/lower signs
correspond to the positive/negative energy modes. Denoting the VEV by $%
\left\langle T_{\mu \nu }\right\rangle =\left\langle 0\right\vert T_{\mu \nu
}\left\vert 0\right\rangle $, with $\left\vert 0\right\rangle $ being the
vacuum state, the mode sum is expressed as%
\begin{equation}
\langle T_{\mu \nu }\rangle =-\frac{i}{4}\sum_{\beta }\sum_{j=+,-}j[\bar{\psi%
}_{\beta }^{(j)}(x)\gamma _{(\mu }\partial _{\nu )}\psi _{\beta
}^{(j)}(x)-(\partial _{(\mu }\bar{\psi}_{\beta }^{(j)}(x))\gamma _{\nu
)}\psi _{\beta }^{(j)}(x)]\ ,  \label{Tmode}
\end{equation}%
where $\gamma _{\mu }=\eta _{\mu \nu }\gamma ^{\nu }$, $\bar{\psi}(x)=\psi
^{\dagger }(x)\gamma ^{0}$ is the Dirac adjoint and the parentheses
including the indices mean symmetrization over those indices. The symbolic
notation $\sum_{\beta }$ stands for the summation over discrete components
of the collective index $\beta $ and the integration over the continuous
ones. The problem under consideration has planar symmetry and it is natural
to take the normal modes corresponding to plane waves.

Taking the chiral representation of the Dirac matrices,%
\begin{equation}
\gamma ^{0}=\left(
\begin{array}{cc}
1 & 0 \\
0 & -1%
\end{array}%
\right) ,\;\gamma ^{l}=\left(
\begin{array}{cc}
0 & \sigma _{l} \\
-\sigma _{l}^{\dagger } & 0%
\end{array}%
\right) ,  \label{gam0}
\end{equation}%
the positive and negative energy wave functions with momentum $\mathbf{k}%
=(k_{1},\ldots ,k_{D})$ and energy $\varepsilon _{\mathbf{k}}=\sqrt{%
k^{2}+m^{2}}$ have the form \cite{Bell14}%
\begin{equation}
\psi _{\beta }^{(\pm )}(x)=\left[ \frac{1+m/\varepsilon _{\mathbf{k}}}{%
2(2\pi )^{p}V_{q}}\right] ^{\frac{1}{2}}e^{i\mathbf{k}\cdot \mathbf{z}\mp
i\varepsilon _{\mathbf{k}}t}\left(
\begin{array}{c}
\left( -\frac{\mathbf{k}\cdot \boldsymbol{\sigma }}{\varepsilon _{\mathbf{k}%
}+m}\right) ^{\frac{1\mp 1}{2}}w_{\chi }^{(\pm )} \\
\left( \frac{\mathbf{k}\cdot \boldsymbol{\sigma }^{\dagger }}{\varepsilon _{%
\mathbf{k}}+m}\right) ^{\frac{1\pm 1}{2}}w_{\chi }^{(\pm )}%
\end{array}%
\right) ,  \label{psibet}
\end{equation}%
where $\mathbf{\sigma }=(\sigma _{1},\ldots ,\sigma _{D})$. Here, the
quantum number $\chi =1,2,\ldots ,N/2$ enumerates the polarization degrees
of freedom, $w_{\chi }^{(\pm )}$ are one-column matrices with $N/2$ rows and
$l$th element $w_{\chi l}^{(\pm )}=\delta _{\chi l}$. We will decompose the
momentum into two parts, $\mathbf{k}=(\mathbf{k}_{p},\mathbf{k}_{q})$, where
$\mathbf{k}_{p}=(k_{1},\ldots ,k_{p})$ and $\mathbf{k}_{q}=(k_{p+1},\ldots
,k_{D})$ are the parts in uncompact and compact subspaces. For the
components $k_{l}$, $l=1,2,\ldots ,p$, one has $-\infty <k_{l}<+\infty $.
The eigenvalues of the components $k_{l}$ along compact dimensions are
quantized by the periodicity conditions (\ref{Qper}):%
\begin{equation}
k_{l}=\frac{2\pi n_{l}+\alpha _{l}}{L_{l}},\;n_{l}=0,\pm 1,\pm 2,\ldots ,
\label{klcomp}
\end{equation}%
for $l=p+1,\ldots ,D$.

The phases in the conditions (\ref{Qper}) can be interpreted in terms of the
magnetic flux for a vector gauge field enclosed by compact dimensions. The
representation described above corresponds to the gauge with zero vector
potential, $(\psi ,A_{\mu })=(\psi ,0)$. Let us pass to a new gauge with the
fields $(\psi ^{\prime },A_{\mu }^{\prime })$, where $A_{\mu }^{\prime }$
has nonzero constant components along compact dimensions: $A_{\mu }^{\prime
}=0$, $\mu =0,1,\ldots ,p$ and $A_{l}^{\prime }=\mathrm{const}$ for $%
l=p+1,\ldots ,D$. The gauge transformation has the form%
\begin{equation*}
\psi ^{\prime }=\psi e^{-ie\lambda },\;A_{\mu }^{\prime }=A_{\mu }+\partial
_{\mu }\lambda ,\;\lambda =b_{\mu }x^{\mu },
\end{equation*}%
with constant $b_{\mu }$ and $e$ being the charge of the Dirac field.
Choosing $b_{\mu }=0$, $\mu =0,1,\ldots ,p$ and $b_{l}=A_{l}^{\prime }$ for $%
l=p+1,\ldots ,D$, for $A_{\mu }=0$ in the new gauge we get $(\psi ^{\prime
},A_{\mu }^{\prime })=(\psi e^{-ieA_{\mu }^{\prime }x^{\mu }},A_{\mu
}^{\prime })$. Taking $A_{l}^{\prime }=\alpha _{l}/(eL_{l})$, we see that
the new field obeys the periodicity condition with $\alpha _{l}^{\prime }=0$%
. Hence, the initial problem with a zero gauge field and quasiperiodicity
condition (\ref{Qper}) is transformed to a gauge with periodic boundary
conditions for the field and with a gauge field having constant components
along compact dimensions. In this interpretation the phases can be expressed
as $\alpha _{l}=-2\pi \Phi _{l}/\Phi _{0}$, where $\Phi _{0}=2\pi /e$ is the
flux quantum and $\Phi _{l}=-A_{l}^{\prime }L_{l}$ is the formal magnetic
flux enclosed by the compact dimension $x^{l}$. That flux takes on real
meaning in models where the space under consideration is embedded in a space
of higher dimension. Examples are the braneworld models and carbon
nanotubes. In the latter case the Dirac field describing the electronic
subsystem of graphene lives in a 2-dimensional space with topology $%
R^{1}\times S^{1}$ and that space is embedded in a 3-dimensional Euclidean
space. The magnetic flux is located inside the tube.

\section{Vacuum energy density and stresses}

\label{sec:EMT}

With the mode functions (\ref{psibet}), the VEV of the energy-momentum
tensor is evaluated by using the formula (\ref{Tmode}). The mode sums for
the energy density and vacuum stresses are transformed to%
\begin{eqnarray}
\langle T_{00}\rangle &=&-\frac{N}{2V_{q}}\int \frac{d\mathbf{k}_{p}}{\left(
2\pi \right) ^{p}}\sum_{\mathbf{n}_{q}\in \mathbf{Z}_{q}}\varepsilon _{%
\mathbf{k}},  \notag \\
\langle T_{\mu \nu }\rangle &=&-\frac{N}{2V_{q}}\int \frac{d\mathbf{k}_{p}}{%
(2\pi )^{p}}\sum_{\mathbf{n}_{q}\in \mathbf{Z}^{q}}\frac{k_{\mu }k_{\nu }}{%
\varepsilon _{\mathbf{k}}},  \label{Tmu}
\end{eqnarray}%
for $\mu ,\nu =1,2,\ldots ,D$, and $\langle T_{0\nu }\rangle =0$. The
component $\langle T_{00}\rangle $ corresponds to the energy density and it
is presented as the sum of the zero-point energies for elementary
oscillators. The expressions (\ref{Tmu}) for the diagonal components have
been considered in \cite{Bell09}. The expressions in (\ref{Tmu}) are
divergent and in \cite{Bell09} two different methods have been used in order
to find the expressions for the renormalized VEVs. The first one is based on
the Abel-Plana summation formula and the second one uses the zeta function
technique \cite{Eliz94,Kirs01}. We will follow the second approach.

For the regularization of the mode sums we introduce the zeta function of a
complex variable $s$ as%
\begin{equation}
\zeta (s)=\frac{1}{V_{q}}\int \frac{d\mathbf{k}_{p}}{\left( 2\pi \right) ^{p}%
}\sum_{\mathbf{n}_{q}\in \mathbf{Z}^{q}}\varepsilon _{\mathbf{k}}^{-2s},
\label{zeta}
\end{equation}%
where the term $\mathbf{n}_{q}=0$ has to be excluded from the sum in the
special case $\alpha _{l}=0$, $l=p+1,\ldots ,D$. After integration over $%
\mathbf{k}_{p}$\ and by using the generalized Chowla-Selberg formula \cite%
{Eliz98,Eliz01} for the resulting series, the zeta function is decomposed as
\cite{Bell09} $\zeta (s)=\zeta _{\mathrm{M}}(s)+\zeta _{\mathrm{t}}(s)$,
where $\zeta _{\mathrm{M}}(s)$ is the corresponding function for the
Minkowski spacetime with trivial topology and the contribution $\zeta _{%
\mathrm{t}}(s)$ is induced by nontrivial topology. Introducing the vectors $%
\boldsymbol{\alpha }_{q}=(\alpha _{p+1},\ldots ,\alpha _{D})$ and $\mathbf{L}%
_{q}=(L_{p+1},\ldots ,L_{D})$ in the compact subspace, the topological part
is expressed as
\begin{equation}
\zeta _{\mathrm{t}}(s)=\frac{2^{1-s}m^{D-2s}}{(2\pi )^{\frac{D}{2}}\Gamma (s)%
}\sideset{}{'}{\sum}_{\mathbf{n}_{q}\in \mathbf{Z}^{q}}\cos (\mathbf{n}%
_{q}\cdot \boldsymbol{\alpha }_{q})f_{\frac{D}{2}-s}(mg(\mathbf{L}_{q},%
\mathbf{n}_{q})),  \label{zetaT}
\end{equation}%
where the prime on the summation sign means that the term $\mathbf{n}_{q}=0$
should be excluded and we have introduced the functions%
\begin{equation}
f_{\nu }(x)=\frac{K_{\nu }(x)}{x^{\nu }},\;g(\mathbf{L}_{q},\mathbf{n}%
_{q})=\left( \sum_{i=p+1}^{D}L_{i}^{2}n_{i}^{2}\right) ^{1/2},  \label{fnu}
\end{equation}%
with $K_{\nu }(x)$ being the modified Bessel function of the second kind.

An alternative representation for the zeta function, convenient in the
asymptotic analysis of the off-diagonal stress, is obtained from (\ref{zetaT}%
) by using the formula (2.40) from \cite{Bell14}. It is given by the formula%
\begin{eqnarray}
\zeta _{\mathrm{t}}(s) &=&\zeta _{p+2,q-2}(s)+\frac{2^{1-s}}{(2\pi )^{\frac{p%
}{2}+1}\Gamma (s)V_{q-2}}\sideset{}{'}{\sum}_{\mathbf{n}_{q}\in \mathbf{Z}%
^{q}}\cos \left( n_{\mu }\alpha _{\mu }\right)  \notag \\
&&\times \cos \left( n_{\nu }\alpha _{\nu }\right) \varepsilon _{\mathbf{n}%
_{q-2}}^{p-2s+2}f_{\frac{p}{2}-s+1}\left( \sqrt{n_{\mu }^{2}L_{\mu
}^{2}+n_{\nu }^{2}L_{\nu }^{2}}\varepsilon _{\mathbf{n}_{q-2}}\right) ,
\label{zetaT2}
\end{eqnarray}%
where $\zeta _{p+2,q-2}(s)$ is the zeta function in the model of topology $%
M^{p+3}\times T^{q-2}$ with decompactified dimensions $x^{\mu }$ and $x^{\nu
}$. Here, $M^{p+3}$ stands for $(p+3)$-dimensional Minkowski spacetime with
trivial topology. The prime on the summation sign in (\ref{zetaT2}) means
that the term $n_{\mu }=n_{\nu }=0$ is excluded from the summation and we
have defined%
\begin{equation}
\varepsilon _{\mathbf{n}_{q-2}}=\left( \sum_{l=p+1,\neq \mu ,\nu
}^{D}k_{l}^{2}+m^{2}\right) ^{1/2}.  \label{epsqm2}
\end{equation}%
The function $\zeta _{p+2,q-2}(s)$ is given by a formula similar to (\ref%
{zetaT}) with the summation over $\mathbf{n}_{q-2}\in \mathbf{Z}^{q-2}$.

The background geometry is flat and the renormalization is reduced to the
subtraction of the Minkowskian VEV. The renormalized energy density is
expressed as $\langle T_{0}^{0}\rangle _{\mathrm{t}}=-N\zeta _{\mathrm{t}%
}(-1/2)/2$. For the diagonal components of the vacuum stresses along
uncompact dimensions one gets (no summation over $\mu $) $\langle T_{\mu
}^{\mu }\rangle _{\mathrm{t}}=\langle T_{0}^{0}\rangle _{\mathrm{t}}$, $\mu
=1,2,\ldots ,p$. The diagonal vacuum stresses in the compact subspace are
found by using the relation (no summation over $\mu $) $\langle T_{\mu
}^{\mu }\rangle _{\mathrm{t}}=\left( L_{\mu }/V_{q}\right) \partial _{L_{\mu
}}\left( V_{q}\langle T_{0}^{0}\rangle _{\mathrm{t}}\right) $, $\mu
=p+1,\ldots ,D$. In this way, from (\ref{zetaT}) for the diagonal component
we find \cite{Bell09} (no summation over $\mu $)
\begin{equation}
\langle T_{\mu }^{\mu }\rangle _{\mathrm{t}}=\frac{Nm^{D+1}}{(2\pi )^{\frac{%
D+1}{2}}}\sideset{}{'}{\sum}_{\mathbf{n}_{q}\in \mathbf{Z}^{q}}\cos (\mathbf{%
n}_{q}\cdot \boldsymbol{\alpha }_{q})F_{(\mu )}(mg(\mathbf{L}_{q},\mathbf{n}%
_{q})),  \label{Tmumu}
\end{equation}%
with the functions%
\begin{equation}
F_{(\mu )}(x)=\left\{
\begin{array}{cc}
f_{\frac{D+1}{2}}(x), & \mu =0,1,2,\ldots ,p \\
f_{\frac{D+1}{2}}(x)-m^{2}L_{\mu }^{2}n_{\mu }^{2}f_{\frac{D+3}{2}}(x), &
\mu =p+1,\ldots ,D%
\end{array}%
\right. .  \label{Fmu}
\end{equation}%
The corresponding expressions in the case of periodic conditions, $\alpha
_{l}=0$, $l=p+1,\ldots ,D$, are obtained from (\ref{Tmumu}) with $\cos (%
\mathbf{n}_{q}\cdot \boldsymbol{\alpha }_{q})=1$. In this case the vacuum
energy density is positive and for the diagonal stresses one has (no
summation over $l$) $\langle T_{l}^{l}\rangle _{\mathrm{t}}<\langle
T_{0}^{0}\rangle _{\mathrm{t}}$, $l=p+1,\ldots ,D$. As it will be shown
below, for periodic conditions the off-diagonal components vanish. By using
the relation
\begin{equation}
x^{2}f_{\nu +1}(x)=f_{\nu -1}(x)+2\nu f_{\nu }(x),  \label{frec}
\end{equation}%
it can be seen that $\sum_{l=1}^{D}\langle T_{l}^{l}\rangle _{\mathrm{t}%
}<\langle T_{0}^{0}\rangle _{\mathrm{t}}$. Hence, the vacuum energy-momentum
tensor for a fermionic field with periodic conditions obey the strong energy
condition. For twisted fields with $\alpha _{l}=\pi $, $l=p+1,\ldots ,D$,
one has $\cos (\mathbf{n}_{q}\cdot \boldsymbol{\alpha }_{q})=(-1)^{n_{p+1}+%
\cdots +n_{D}}$.

The result (\ref{Tmumu}) shows that the vacuum stresses in the uncompact
subspace are equal to the energy density, $\langle T_{\mu }^{\mu }\rangle _{%
\mathrm{t}}=\langle T_{0}^{0}\rangle _{\mathrm{t}}$ , $\mu =1,\ldots ,p$ (no
summation over $\mu $). Of course, this is a consequence of the Lorentz
invariance in that subspace. By taking into account that for the vacuum
effective pressure along the direction $x^{\mu }$ one has $P_{\mu }=-\langle
T_{\mu }^{\mu }\rangle _{\mathrm{t}}$, we see that the equation of state for
the vacuum in the uncompact subspace is of the cosmological constant type.
The models with the topological Casimir energy as the source of the
accelerated expansion are based on this property.

For a massless fermionic field the general result (\ref{Tmumu}) is reduced
to (no summation over $\mu $)
\begin{equation}
\left\langle T_{\mu }^{\mu }\right\rangle _{\mathrm{t}}=\frac{N}{2\pi ^{%
\frac{D+1}{2}}}\Gamma \left( \frac{D+1}{2}\right) \sideset{}{'}{\sum}_{%
\mathbf{n}_{q}\in \mathbf{Z}^{q}}\frac{\cos (\mathbf{n}_{q}\cdot \boldsymbol{%
\alpha }_{q})}{g^{D+1}(\mathbf{L}_{q},\mathbf{n}_{q})}F_{(\mu )}^{(0)}(%
\mathbf{L}_{q},\mathbf{n}_{q}),  \label{Tmumum0}
\end{equation}%
where $F_{(\mu )}^{(0)}(\mathbf{L}_{q},\mathbf{n}_{q})=1$ for $\mu
=0,1,2,\ldots ,p$, and%
\begin{equation}
F_{(\mu )}^{(0)}(\mathbf{L}_{q},\mathbf{n}_{q})=1-\frac{\left( D+1\right)
n_{\mu }^{2}L_{\mu }^{2}}{g^{2}(\mathbf{L}_{q},\mathbf{n}_{q})},
\label{Fmu0}
\end{equation}%
for $\mu =p+1,\ldots ,D$. In this special case the vacuum energy-momentum
tensor is traceless, $\left\langle T_{\mu }^{\mu }\right\rangle _{\mathrm{t}%
}=0$.

Here we are interested in the off-diagonal components. For $\mu
=0,1,2,\ldots ,p$ and $\nu \neq \mu $ one gets $\langle T_{\mu \nu }\rangle
_{\mathrm{t}}=0$. The possible nonzero components $\langle T_{\mu \nu
}\rangle _{\mathrm{t}}$ correspond to $\mu ,\nu =p+1,\ldots ,D$. In order to
use the zeta function, we note that the following relation takes place%
\begin{equation}
\frac{k_{\nu }k_{\mu }}{\varepsilon _{\mathbf{k}}}=\frac{L_{\mu }L_{\nu }}{%
3\left( 2\pi \right) ^{2}}\partial _{\alpha _{\mu }}\partial _{\alpha _{\nu
}}\varepsilon _{\mathbf{k}}^{3}.  \label{Rel1}
\end{equation}%
This allows to write the off-diagonal components in the form%
\begin{equation}
\langle T_{\mu \nu }\rangle =-\frac{NL_{\mu }L_{\nu }}{6\left( 2\pi \right)
^{2}}\partial _{\alpha _{\mu }}\partial _{\alpha _{\nu }}\frac{1}{V_{q}}\int
\frac{d\mathbf{k}_{p}}{(2\pi )^{p}}\sum_{\mathbf{n}_{q}\in \mathbf{Z}%
^{q}}\varepsilon _{\mathbf{k}}^{3}=-\frac{NL_{\mu }L_{\nu }}{6\left( 2\pi
\right) ^{2}}\partial _{\alpha _{\mu }}\partial _{\alpha _{\nu }}\zeta
\left( -\frac{3}{2}\right) .  \label{Tmunu}
\end{equation}%
By using the formula (\ref{zetaT}), for the topological part we get%
\begin{equation}
\langle T_{\mu \nu }\rangle _{\mathrm{t}}=\frac{m^{D+3}NL_{\mu }L_{\nu }}{%
(2\pi )^{\frac{D+1}{2}}}\sideset{}{'}{\sum}_{\mathbf{n}_{q}\in \mathbf{Z}%
^{q}}n_{\mu }n_{\nu }\cos (\mathbf{n}_{q}\cdot \boldsymbol{\alpha }_{q})f_{%
\frac{D+3}{2}}(mg(\mathbf{L}_{q},\mathbf{n}_{q})).  \label{Tmunu1}
\end{equation}%
Note that in this representation we can make the replacement%
\begin{equation}
\cos (\mathbf{n}_{q}\cdot \boldsymbol{\alpha }_{q})\rightarrow \sin \left(
n_{\mu }\alpha _{\mu }\right) \sin \left( n_{\nu }\alpha _{\nu }\right) \cos
(\mathbf{n}_{q-2}\cdot \boldsymbol{\alpha }_{q-2}),  \label{Repl1}
\end{equation}%
where $\mathbf{n}_{q-2}\cdot \boldsymbol{\alpha }_{q-2}=\sum_{l\neq \mu ,\nu
}n_{l}\alpha _{l}$. This replacement explicitly shows that the off-diagonal
component $\langle T_{\mu \nu }\rangle _{\mathrm{t}}$ is an even periodic
function of the phases $\alpha _{l}$, $l\neq \mu ,\nu $, with the period
equal to $2\pi $, and an odd periodic function of $\alpha _{\mu }$ and $%
\alpha _{\nu }$ with the same period. Hence, without the loss of generality,
we can assume that $|\alpha _{\mu }|\leq \pi $. For a massless field, by
taking into account that $f_{\nu }(x)=2^{\nu -1}\Gamma (\nu )x^{-2\nu }$ for
$x\ll 1$, one obtains%
\begin{equation}
\langle T_{\mu \nu }\rangle _{\mathrm{t}}=\frac{NL_{\mu }L_{\nu }}{\pi ^{%
\frac{D+1}{2}}}\Gamma \left( \frac{D+3}{2}\right) \sideset{}{'}{\sum}_{%
\mathbf{n}_{q}\in \mathbf{Z}^{q}}n_{\mu }n_{\nu }\frac{\cos (\mathbf{n}%
_{q}\cdot \boldsymbol{\alpha }_{q})}{g^{D+3}(\mathbf{L}_{q},\mathbf{n}_{q})}.
\label{Tmunum0}
\end{equation}

An equivalent expression for the off-diagonal stresses is obtained by using
the representation (\ref{zetaT2}) for the zeta function in (\ref{Tmunu}).
The first term in the right-hand side of (\ref{zetaT2}) does not depend on $%
\alpha _{\mu }$ and $\alpha _{\nu }$ and, hence, it does not contribute to
the stress. The following expression is obtained:%
\begin{eqnarray}
\langle T_{\mu \nu }\rangle _{\mathrm{t}} &=&-\frac{NL_{\mu }^{2}L_{\nu }^{2}%
}{(2\pi )^{\frac{p+3}{2}}V_{q}}\sideset{}{'}{\sum}_{\mathbf{n}_{q}\in
\mathbf{Z}^{q}}n_{\mu }n_{\nu }\sin \left( n_{\mu }\alpha _{\mu }\right)
\notag \\
&&\times \sin \left( n_{\nu }\alpha _{\nu }\right) \varepsilon _{\mathbf{n}%
_{q-2}}^{p+5}f_{\frac{p+5}{2}}\left( \sqrt{n_{\mu }^{2}L_{\mu }^{2}+n_{\nu
}^{2}L_{\nu }^{2}}\varepsilon _{\mathbf{n}_{q-2}}\right) .  \label{Tmunu3}
\end{eqnarray}%
For $q=2$ ($p=D-2$) one has $\varepsilon _{\mathbf{n}_{q-2}}=m$ and the
formula is reduced to%
\begin{equation}
\langle T_{\mu \nu }\rangle _{\mathrm{t}}=-\frac{Nm^{D+3}L_{\mu }L_{\nu }}{%
(2\pi )^{\frac{D+1}{2}}}\sideset{}{'}{\sum}_{n_{\mu },n_{\nu }=-\infty
}^{+\infty }n_{\mu }n_{\nu }\sin \left( n_{\mu }\alpha _{\mu }\right) \sin
\left( n_{\nu }\alpha _{\nu }\right) f_{\frac{D+3}{2}}\left( m\sqrt{n_{\mu
}^{2}L_{\mu }^{2}+n_{\nu }^{2}L_{\nu }^{2}}\right) .  \label{Tmunuq2}
\end{equation}%
In the special case under consideration this coincides with (\ref{Tmunu1}).

The special case of the results corresponding to $D=2$ describes the
properties of the ground state for the electronic subsystem in graphene
nanotubes and nanoloops (toroidal nanotubes) described by the effective
Dirac model. For nanotubes one has $(p,q)=(1,1)$ and for nanoloops $%
(p,q)=(0,2)$. For metallic nanotubes and in the absence of the threading
magnetic flux the phase along the periodic condition is zero, $\alpha _{l}=0$%
. Depending on the chiral vector in semiconductor nanotubes two values of
the phases are realized with $\alpha _{l}=\pi /3$ and $\alpha _{l}=\pi /3$.
The corresponding analysis for the diagonal components of the ground state
energy-momentum tensor can be found in \cite{Bell09}. The off-diagonal
component for nanoloops is obtained from (\ref{Tmunuq2}) with $D=2$ and $N=2$%
. In this special case one has $f_{5/2}=\sqrt{\pi /2}%
x^{-5}e^{-x}(3+3x+x^{2}) $.

\section{Asymptotic analysis and numerical results}

\label{sec:Asnum}

Let us consider some asymptotics of general formulae. For large values of
the length of the compact dimension $z^{l}$, $l\neq \mu ,\nu $, $L_{l}\gg
L_{\mu },L_{\nu }$, the dominant contribution in (\ref{Tmunu1}) comes from
the term with $n_{l}=0$ and the leading order term coincides with the
off-diagonal stress in the model where the coordinate $z^{l}$ is
decompactified. In the opposite limit $L_{l}\ll L_{\mu },L_{\nu }$, it is
more convenient to use the the representation (\ref{Tmunu3}). The behavior
of the stress is essentially different depending on whether the phase $%
\alpha _{l}$ is zero or not. For $\alpha _{l}=0$ the main contribution to
the VEV $\langle T_{\mu \nu }\rangle _{\mathrm{t}}$ comes from the modes
with $n_{l}=0$. To the leading order we get%
\begin{equation}
\langle T_{\mu \nu }\rangle _{\mathrm{t}}\approx \frac{N\langle T_{\mu \nu
}\rangle _{\mathrm{t}}^{(M^{p+1}\times T^{q-1})}}{N_{D-1}L_{l}},
\label{Tmunuap1}
\end{equation}%
where $N_{D-1}=2^{\left[ \frac{D}{2}\right] }$ and $\langle T_{\mu \nu
}\rangle _{\mathrm{t}}^{(M^{p+1}\times T^{q-1})}$ is the corresponding VEV
in $D$-dimensional spacetime with topology $M^{p+1}\times T^{q-1}$ which is
obtained from the geometry described by (\ref{ds}) excluding the compact
dimension $x^{l}$. For $\alpha _{l}\neq 0$ and assuming that $|\alpha
_{l}|<\pi $, again, the dominant contribution give the modes $n_{l}=0$. The
argument of the function $f_{(p+5)/2}(x)$ is large and we can use the
corresponding asymptotic of the modified Bessel function. This shows that in
the limit under consideration the off-diagonal stress $\langle T_{\mu \nu
}\rangle _{\mathrm{t}}$ is suppressed by the factor $\exp (-|\alpha _{l}|%
\sqrt{L_{\mu }^{2}+L_{\nu }^{2}}/L_{l})$.

For large values of the lengths $L_{\mu }$ and $L_{\nu }$ compared to the
other length scales $1/m$ and $L_{l}$, $l\neq \mu ,\nu $, by using (\ref%
{Tmunu3}) we can see that the topological contribution $\langle T_{\mu \nu
}\rangle _{\mathrm{t}}$ is exponentially suppressed by the factor $\exp
(-\varepsilon _{(0)}\sqrt{L_{\mu }^{2}+L_{\nu }^{2}})$, where $\varepsilon
_{(0)}=\varepsilon _{\mathbf{n}_{q-2}}|_{\mathbf{n}_{q-2}=0}$ and $0<|\alpha
_{l}|<\pi $. For small values of $L_{\mu }$ and $L_{\nu }$ the dominant
contribution in (\ref{Tmunu3}) comes from large values of $|n_{l}|$ and we
can replace the corresponding summations by the integration in accordance
with%
\begin{equation}
\frac{L_{\mu }L_{\nu }}{V_{q}}\sideset{}{'}{\sum}_{\mathbf{n}_{q-2}\in
\mathbf{Z}^{q-2}}\varepsilon _{\mathbf{n}_{q-2}}^{p+5}f_{\frac{p+5}{2}%
}\left( b\varepsilon _{\mathbf{n}_{q-2}}\right) \rightarrow \int \frac{%
d^{q-2}\mathbf{y}}{\left( 2\pi \right) ^{q-2}}\left( \mathbf{y}%
^{2}+m^{2}\right) ^{(p+5)/2}f_{\frac{p+5}{2}}\left( b\sqrt{\mathbf{y}%
^{2}+m^{2}}\right) ,  \label{Repl}
\end{equation}%
where $b=\sqrt{n_{\mu }^{2}L_{\mu }^{2}+n_{\nu }^{2}L_{\nu }^{2}}$ and $%
\mathbf{y}=(y_{1},\ldots ,y_{q-2})$ with $-\infty <y_{l}<+\infty $. After
integration over the angular part, the integral over $|\mathbf{y}|$ is
evaluated by using the formula from \cite{Grad07}. In this way it can be
seen that the leading term in the expansion of $\langle T_{\mu \nu }\rangle
_{\mathrm{t}}$ coincides with (\ref{Tmunuq2}). Additionally assuming that $m%
\sqrt{L_{\mu }^{2}+L_{\nu }^{2}}\ll 1$, in the leading approximation we get%
\begin{equation}
\langle T_{\mu \nu }\rangle _{\mathrm{t}}\approx -\frac{NL_{\mu }L_{\nu }}{%
\pi ^{\frac{D+1}{2}}}\Gamma \left( \frac{D+3}{2}\right) \sideset{}{'}{\sum}%
_{n_{\mu },n_{\nu }=-\infty }^{+\infty }n_{\mu }n_{\nu }\frac{\sin \left(
n_{\mu }\alpha _{\mu }\right) \sin \left( n_{\nu }\alpha _{\nu }\right) }{%
\left( n_{\mu }^{2}L_{\mu }^{2}+n_{\nu }^{2}L_{\nu }^{2}\right) ^{\frac{D+3}{%
2}}}.  \label{Tmunuap2}
\end{equation}%
Note that the right-hand side presents the off-diagonal component of the
vacuum energy-momentum tensor for a massless fermionic field in the model $%
(p,q)=(D-2,2)$ with compact dimensions $x^{\mu }$ and $x^{\nu }$.

\begin{figure}[tbph]
\begin{center}
\begin{tabular}{cc}
\epsfig{figure=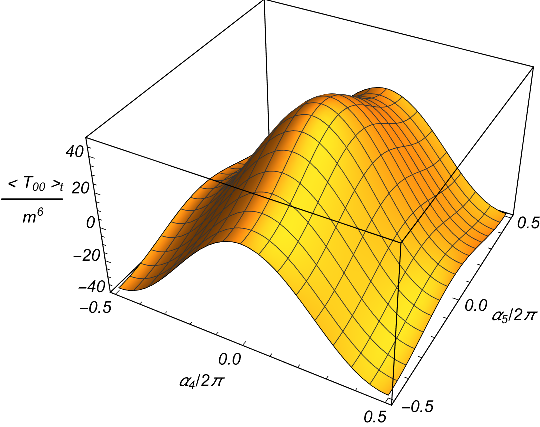,width=7.5cm,height=6cm} & \quad %
\epsfig{figure=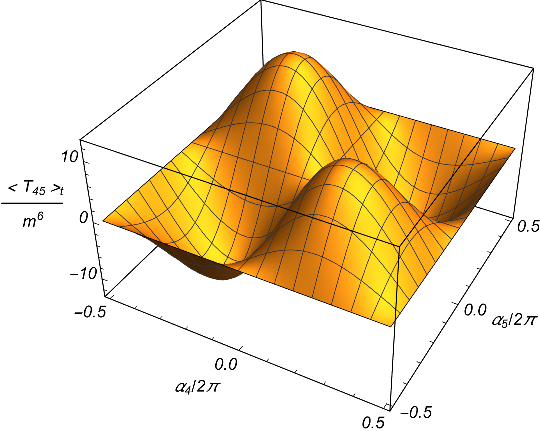,width=7.5cm,height=6cm}%
\end{tabular}%
\end{center}
\caption{The expectation values of the vacuum energy density and
off-diagonal stress on the phases of the periodicity conditions in the model
$(p,q)=(3,2)$ with $mL_{4}=0.5$, $mL_{5}=0.6$. }
\label{fig1}
\end{figure}

We will present the numerical analysis for the $D=5$ model with two compact
dimensions $x^{4}$ and $x^{5}$. This corresponds to the set $(p,q)=(3,2)$.
By taking into account that in \cite{Bell09} the numerical results for the
energy density and diagonal stresses are given for the model $D=4$ with a
single compact dimension, the analysis will be given for those quantities as
well. We start from the dependence of the expectation values on the phases $%
\alpha _{4}$ and $\alpha _{5}$. Figure \ref{fig1} presents that dependence
of the energy density (left panel) and off-diagonal stress $\langle
T_{45}\rangle _{\mathrm{t}}$ (right panel) for $mL_{4}=0.5$ and $mL_{6}=0.6$.

\begin{figure}[tbph]
\begin{center}
\begin{tabular}{cc}
\epsfig{figure=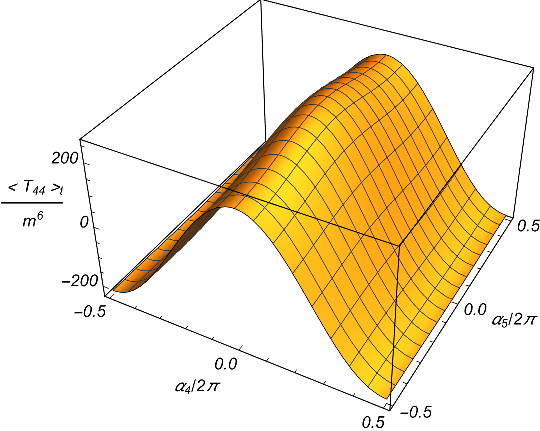,width=7.5cm,height=6cm} & \quad %
\epsfig{figure=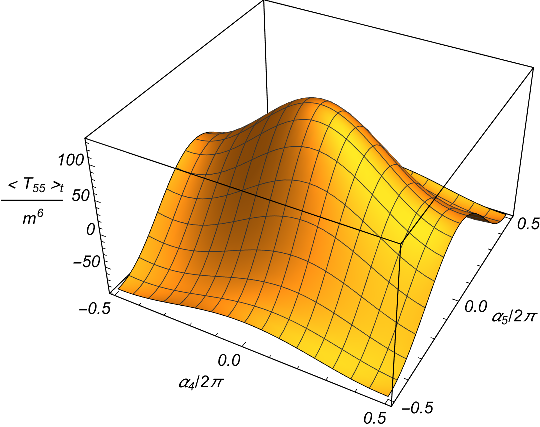,width=7.5cm,height=6cm}%
\end{tabular}%
\end{center}
\caption{The same as in figure \protect\ref{fig1} for the stresses along the
compact dimensions $x^{4}$ (left panel) and $x^{5}$ (right panel).}
\label{fig2}
\end{figure}

The corresponding results for the diagonal stresses along compact dimensions
are given in figure \ref{fig2}. As already mentioned above, the energy
density and the diagonal stresses are even periodic functions of $\alpha
_{4} $ and $\alpha _{5}$, whereas the off-diagonal component is an odd
periodic function of those phases. Depending on the specific values of the
phases, all the components can be either positive or negative. The energy
density $\langle T_{00}\rangle _{\mathrm{t}}$ and the vacuum pressures $%
\langle T_{44}\rangle _{\mathrm{t}}$ and $\langle T_{55}\rangle _{\mathrm{t}%
} $ are positive for the values of the phases near $(\alpha _{4},\alpha
_{5})=(0,0)$ (periodic conditions) and negative near $(\alpha _{4},\alpha
_{5})=(\pi ,\pi )$ (antiperiodic conditions).

\begin{figure}[tbph]
\begin{center}
\begin{tabular}{cc}
\epsfig{figure=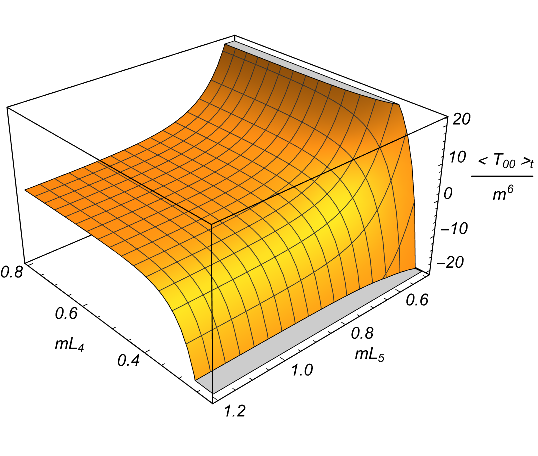,width=7.5cm,height=6cm} & \quad %
\epsfig{figure=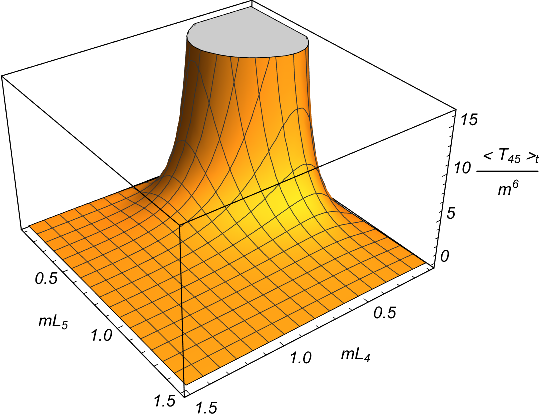,width=7.5cm,height=6cm}%
\end{tabular}%
\end{center}
\caption{The expectation values of the vacuum energy density and
off-diagonal stress on the length of compact dimensions in the model $%
(p,q)=(3,2)$. For the left panel we have taken $\protect\alpha _{4}=\protect%
\pi /2$, $\protect\alpha _{5}=0$ and for the right panel $\protect\alpha %
_{4}=\protect\pi /2$, $\protect\alpha _{5}=-0.6\protect\pi $. }
\label{fig3}
\end{figure}

\begin{figure}[tbph]
\begin{center}
\begin{tabular}{cc}
\epsfig{figure=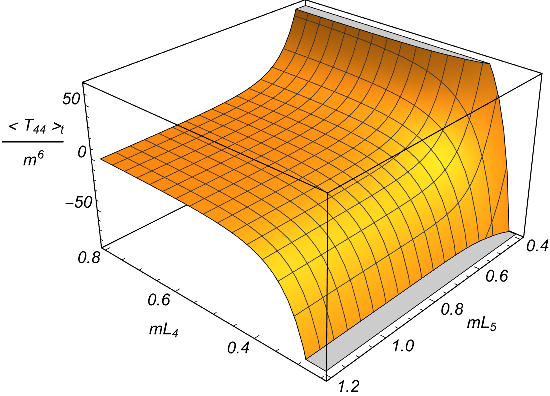,width=7.5cm,height=6cm} & \quad %
\epsfig{figure=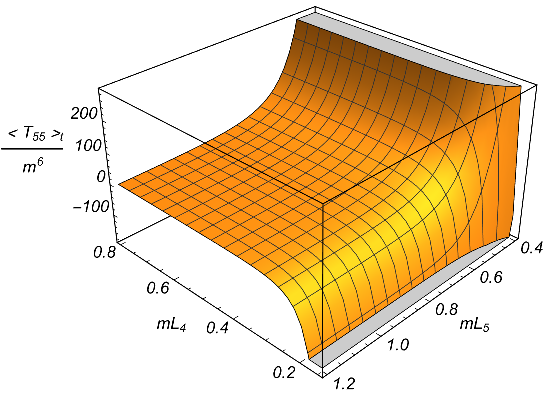,width=7.5cm,height=6cm}%
\end{tabular}%
\end{center}
\caption{The vacuum stresses along the compact dimensions $x^{4}$ (left
panel) and $x^{5}$ (right panel) versus the lengths of those dimensions. The
graphs are plotted for $\protect\alpha _{4}/2\protect\pi =0.25$, $\protect%
\alpha _{5}/2\protect\pi =0$.}
\label{fig4}
\end{figure}

The dependence of the VEVs of the components for the energy-momentum tensor
on the lengths of compact dimensions is presented in figures \ref{fig3}
(energy density and off-diagonal component) and \ref{fig4} (stresses along
compact dimensions). For the diagonal components we have taken the phases $%
\alpha _{4}=\pi /2$ and $\alpha _{5}=0$ and the off-diagonal component is
plotted for $\alpha _{4}=\pi /2$, $\alpha _{5}=-0.6\pi $.


\begin{figure}[tbph]
\begin{center}
\epsfig{figure=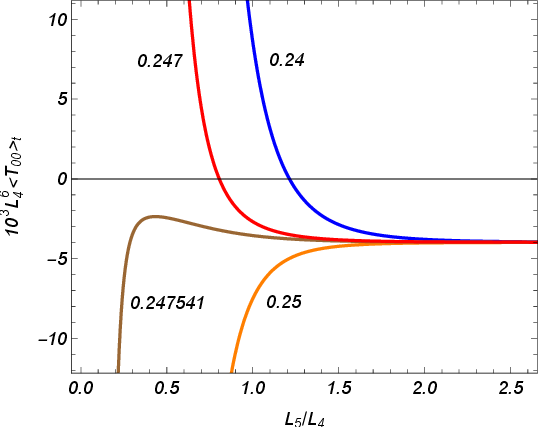,width=8.5cm,height=7cm}
\end{center}
\caption{The energy density for a massless fermionic field in the model $%
(p,q)=(3,2)$ as a function of the ratio $L_{5}/L_{4}$ for fixed value $%
\protect\alpha _{4}=\protect\pi /2$. The numbers near the graphs are the
values of $\protect\alpha _{5}/2\protect\pi $. }
\label{fig5}
\end{figure}

From the given graphs, one can get the impression that the energy density is
a monotonic function of the lengths of the compact dimensions. However, this
is not the case even for a massless field. In order to demonstrate that and
by taking into account that the VEVs for a massless field approximate the
results for massive fields in the limit of small values of the lengths of
compact dimensions, in figure \ref{fig5} we have plotted the dimensionless
quantity $L_{4}^{6}\langle T_{00}\rangle _{\mathrm{t}}$ as a function of the
ratio $L_{5}/L_{4}$. The corresponding expression is given by the right-hand
side of (\ref{Tmunuap2}). The graphs are plotted for $\alpha _{4}=\pi /2$
and the numbers near the curves are the values of the ratio $\alpha
_{5}/2\pi $. For large values of $L_{5}/L_{4}$ all the curves tend to the
corresponding result for the energy density in the model where the direction
$x^{5}$ is decompactified ($L_{5}\rightarrow \infty $).

\section{Conclusions}

\label{sec:Conc}

Continuing the investigations started in \cite{Bell09} we have studied the
effects of nontrivial topology on the local characteristics of the fermionic
vacuum. A toroidal compactification of a part of spatial dimensions in $(D+1)
$-dimensional flat spacetime is considered. In addition to the diagonal
components, studied in \cite{Bell09}, the vacuum energy-momentum tensor has
an off-diagonal components having indices along compact dimensions. Those
components vanish for periodic ($\alpha _{l}=0$) and antiperiodic ($\alpha
_{l}=\pi $) conditions. In the first case the vacuum energy-momentum tensor
for a fermionic field obeys the strong energy condition. For general values
of the phases that is not the case. The phases in the periodicity conditions
can be interpreted in terms of magnetic fluxes enclosed by compact
dimensions. The VEVs are periodic functions of magnetic fluxes with the
period of flux quantum. The diagonal components are even functions of the
phases $\alpha _{l}$. The off-diagonal component $\langle T_{\mu \nu
}\rangle _{\mathrm{t}}$, $\mu \neq \nu $, $\mu ,\nu =p+1,\ldots ,D$, is an
even function of $\alpha _{l}$ with $l\neq \mu ,\nu $, and odd function of
the phases $\alpha _{\mu }$ and $\alpha _{\nu }$. The vacuum stresses in the
uncompact subspace are isotropic and the corresponding equation of state is
of the cosmological constant type. Depending on the values of the phases the
components of the vacuum energy-momentum tensor can be either positive or
negative. For small values of the lengths $L_{\mu }$ and $L_{\nu }$, the
off-diagonal component is approximated by the corresponding result for a
massless field in the model with $q=2$ and compact subspace $(x^{\mu
},x^{\nu })$ (see (\ref{Tmunuap2})). The numerical analysis of the obtained
results is presented for the $D=5$ with $(p,q)=(3,2)$.

We have considered the effects of the nontrivial topology on the local
properties of the fermionic vacuum. In the presence of boundaries additional
contributions are induced in the VEVs of physical observables (the
boundary-induced Casimir effect). The effects of two planar boundaries with
the bag boundary conditions on the Dirac field in the geometry under
consideration have been discussed in \cite{Bell09b,Eliz11}. The results in
the special case of 2-dimensional space are applied to finite length carbon
nanotubes. The fermionic condensate and the VEV of the energy-momentum
tensor in toroidally compactified de Sitter spacetime are studied in \cite%
{Saha08}.

\section*{Acknowledgments}

The work was supported by the grant No. 21AG-1C047 of the Higher Education
and Science Committee of the Ministry of Education, Science, Culture and
Sport RA.

\end{document}